\documentclass{lhep}       

\journal{LHEP}%Letters in High Energy Physics}
\vol{xx}
\jyear{2022}
\pages{xxx} 

\publishedtrue
\received{October 2022}
\published{2022}

\def\be{\begin{equation}}
\def\ee{\end{equation}}
\def\bea{\begin{eqnarray}}
\def\eea{\end{eqnarray}}

\begin{document}

\title{Potential for definitive discovery of a 70 GeV dark matter WIMP \\ with only second-order gauge couplings}

\author{Bailey Tallman, Alexandra Boone, Adhithya Vijayakumar, Fiona Lopez, Samuel Apata, Jehu Martinez, and Roland Allen}
\address{Physics and Astronomy Department, Texas A\&M University, College Station, Texas 77843, USA}

\begin{abstract}
As astronomical observations and their interpretation improve, the case for cold dark matter (CDM) becomes increasingly persuasive. A particularly appealing version of CDM is a weakly interacting massive particle (WIMP) with a mass near the electroweak scale, which can naturally have the observed relic abundance after annihilation in the early universe. But in order for a WIMP to be consistent with the currently stringent experimental constraints it must have relatively small cross-sections for indirect, direct, and collider detection.  Using our calculations and estimates of these cross-sections, we discuss the potential for discovery of a recently proposed dark matter WIMP which has a mass of about 70 GeV/c$^2$ and  only second-order couplings to W and Z bosons. There is evidence that indirect detection may already have been achieved, since analyses of the gamma rays detected by Fermi-LAT and the antiprotons observed by AMS-02 are consistent with 70 GeV dark matter having our calculated $\langle \sigma_{ann} v \rangle \approx 1.2 \times 10^{-26} $ cm$^3$/s. The estimated sensitivities for LZ and XENONnT indicate that these experiments may achieve direct detection within the next few years, since we estimate the relevant cross-section to be slightly above $10^{-48}$ cm$^2$. Other experiments such as PandaX, SuperCDMS, and especially DARWIN should be able to confirm on a longer time scale. The high-luminosity LHC might achieve collider detection within about 15 years, since we estimate a collider cross-section slightly below 1 femtobarn. Definitive confirmation should come from still more powerful planned collider experiments (such as a future circular collider) within 15-35 years. 
\end{abstract}

\maketitle

\begin{keyword}
dark matter
%\doi{10.2018/LHEP000001}
\end{keyword}

\hspace{1cm}

There are many aspects of the dark matter problem~\cite{Silk,Mambrini} and a vast number of dark matter candidates~\cite{Snowmass,pdg}, with masses and couplings spanning many orders of magnitude. The cold dark matter (CDM) paradigm has, however, become increasingly compelling during the past quarter century, because of the growing sophistication of astronomical observations and their interpretation~\cite{pdg,DES-Wechsler}. A particularly appealing version of CDM continues to be weakly interacting massive particles (WIMPs), since a weakly interacting particle with a mass near the electroweak scale can naturally emerge from the early universe with about the observed relic abundance. 
\begin{figure}[!tbp]
\begin{center}
\resizebox{1.00\columnwidth}{!}{\includegraphics{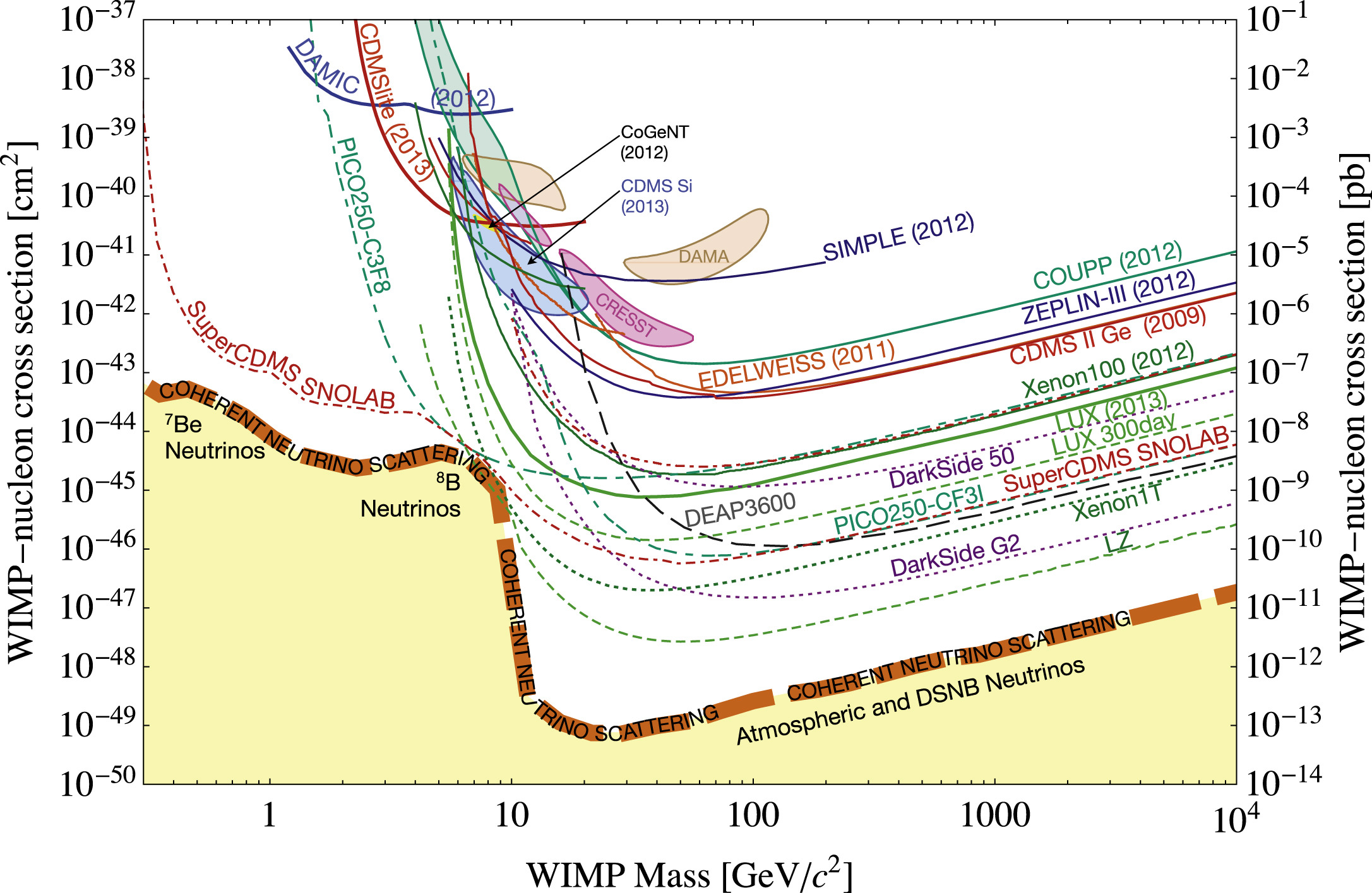}}
\end{center}
\caption{Reach of previous direct detection experiments. From Ref.~\cite{Strigari-2014}, used with permission. The present dark matter candidate has couplings to only W and Z bosons, and these are only second-order. It consequently has only a small cross-section for scattering off atomic nuclei, estimated to be slightly above $10^{-48}$cm$^2$ in the case of Xe~\cite{DM2022a}, so it lies below the sensitivities of earlier experiments. With a mass of about 70 GeV/c$^2$, it should barely be detectable by the LZ and XENONnT experiments, both of which estimate a reach down to about $1.4 \times 10^{-48}$cm$^2$ for a dark matter particle with a mass $\sim 50$ GeV/c$^2$. The current and projected sensitivities of LZ and XENONnT, shown in Figs.~\ref{LZ_July} - \ref{XENON_ty}, demonstrate the grounds for this prediction in more detail. }
\label{Strigari-2014}
\end{figure}

There are, however, stringent limits on the cross-sections for direct, indirect, and collider detection. Figure~\ref{Strigari-2014} shows the remarkable sensitivity achieved in direct detection experiments during the past few decades~\cite{Strigari-2014}, which demonstrates that a viable dark matter candidate must have a very small cross-section for scattering off an atomic nucleus. 

\begin{figure}[!tbp]
\begin{center}
\resizebox{1.00\columnwidth}{!}{\includegraphics{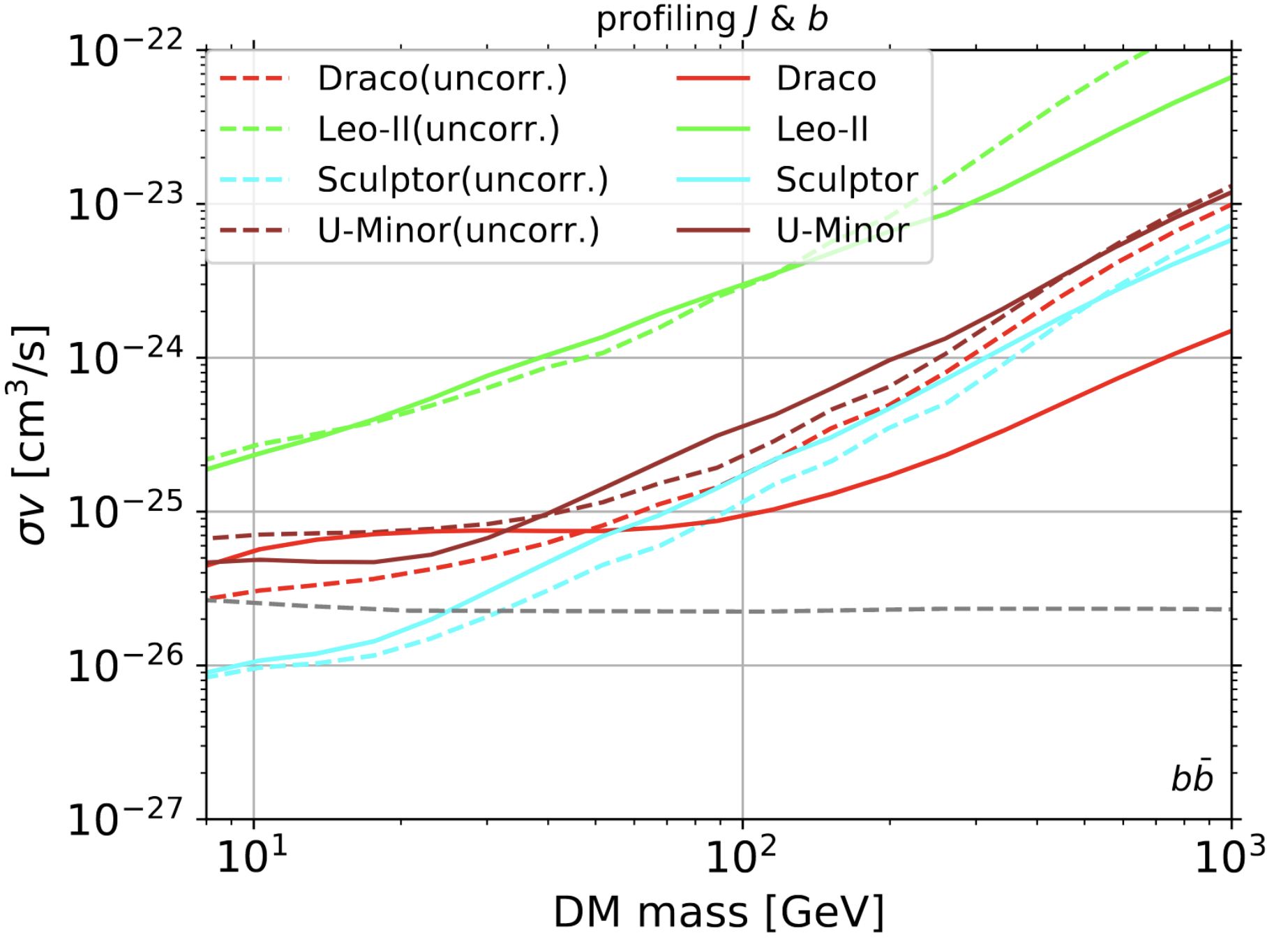}}
\end{center}
\caption{Upper bounds on $\langle \sigma_{ann} v \rangle $ from Fermi-LAT gamma-ray observations of dwarf spheroidal galaxies near the Milky Way. The solid and dashed curves are two limiting cases which ``should bracket somewhat the real energy correlation''. The dashed gray line indicates the thermal relic cross section inferred for generic WIMP models~\cite{Steigman}. From Ref.~\cite{Calore-2020}, used with permission.}
\label{Calore-2020}
\end{figure}

As can be seen in Fig.~\ref{Calore-2020}, there are also strong bounds on the cross-section for annihilation in the present universe, determined by observations of dwarf spheroidal galaxies~\cite{Calore-2020}. 

Finally, the hopes for collider detection at the LHC have not been realized, and strong limits have been placed on new particles of any kind, including dark matter particles~\cite{collider,pdg2}.

Here we will focus on the potential for detection of a new dark matter particle which is consistent with all experimental and observational limits, and which additionally appears to be the only viable candidate with a well-defined mass and well-defined couplings~\cite{DM2021a,DM2021b,DM2022a}. Since there are no free parameters, it is possible to determine the cross-sections for indirect, direct, and collider detection, providing clean experimental tests of the theory. 

This candidate is a WIMP with a mass of about 70 GeV/c$^2$ and an annihilation cross section in the present universe given by $\langle \sigma_{ann} v \rangle \approx 1.2 \times 10^{-26} $ cm$^3$/s, according to the calculations described below, if it is assumed to constitute 100\% of the dark matter. It should be mentioned, however, that the present theory also predicts supersymmetry (susy) at some energy scale, and that the lightest superpartner~\cite{Silk,susy-DM-1996,Baer-Tata,Kane,snowmass-2013,Baer-Roszkowski-2015,Baer-Barger-2016,Baer-Barger-2018,Roszkowski-2018,Baer-Barger-2020,Tata-2020} can be a subdominant component in a multicomponent scenario.

\begin{figure}[t]
\begin{center}
\resizebox{1.00\columnwidth}{!}{\includegraphics{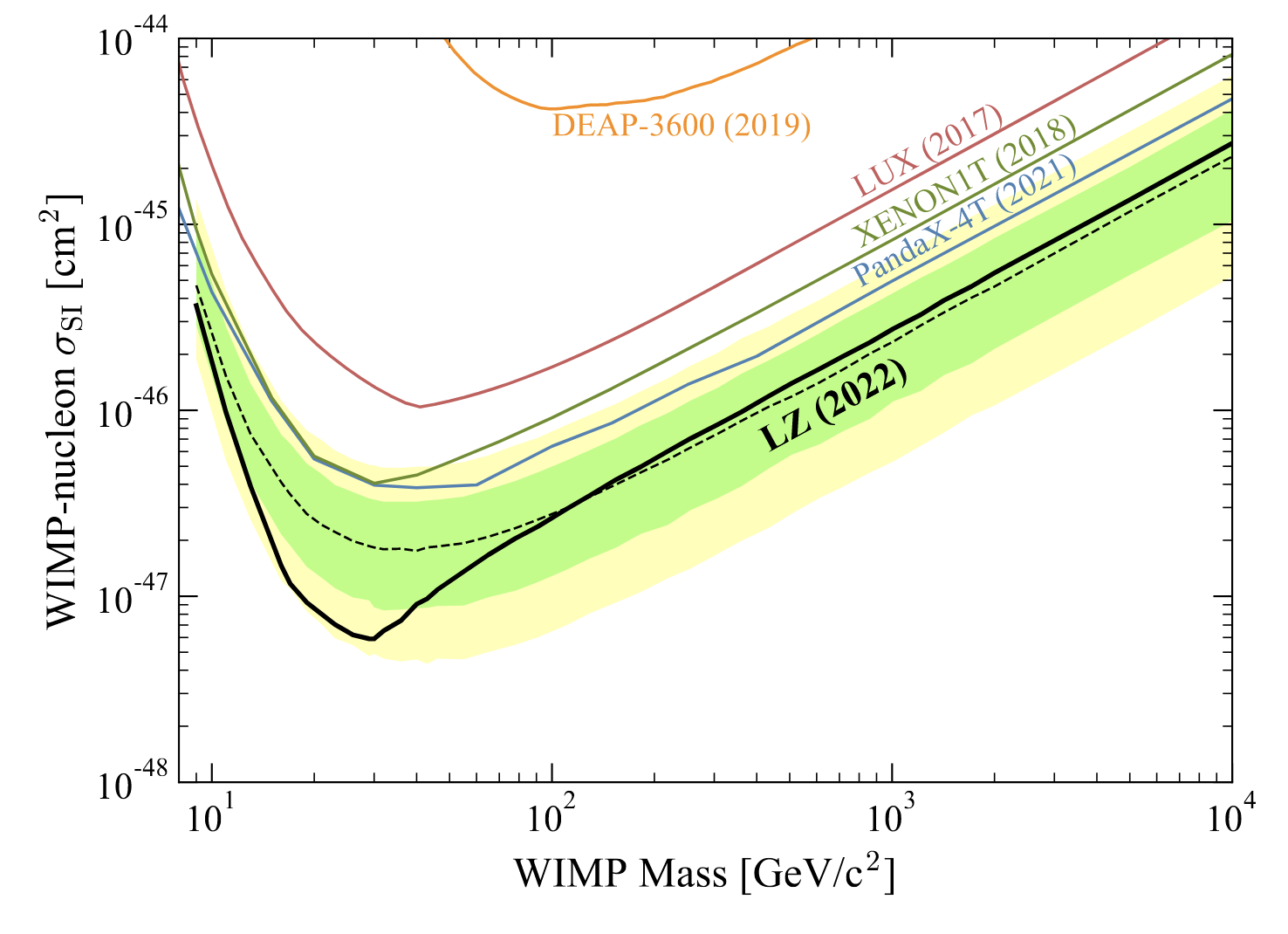}}
\end{center}
\caption{Reach of LZ in July 2022. From Ref.~\cite{LZ-2022}, used with permission.}
\label{LZ_July}
\end{figure}
\begin{figure}[t]
\begin{center}
\resizebox{1.00\columnwidth}{!}{\includegraphics{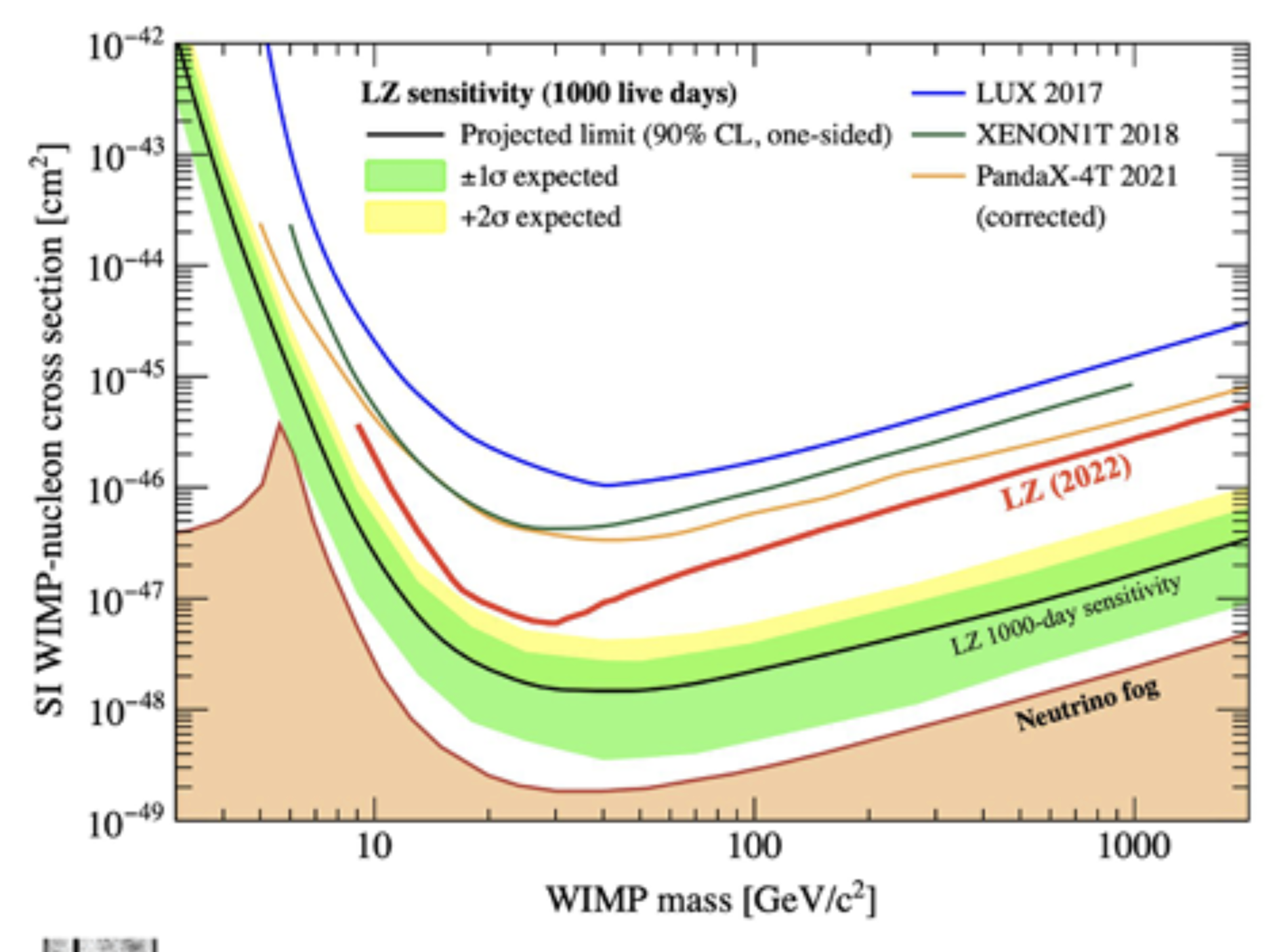}}
\end{center}
\caption{Reach of LZ with 1000 days of data. From Ref.~\cite{LZ-1000}, used with permission.}
\label{LZ_1000}
\end{figure}\begin{figure}[t]
\begin{center}
\resizebox{1.00\columnwidth}{!}{\includegraphics{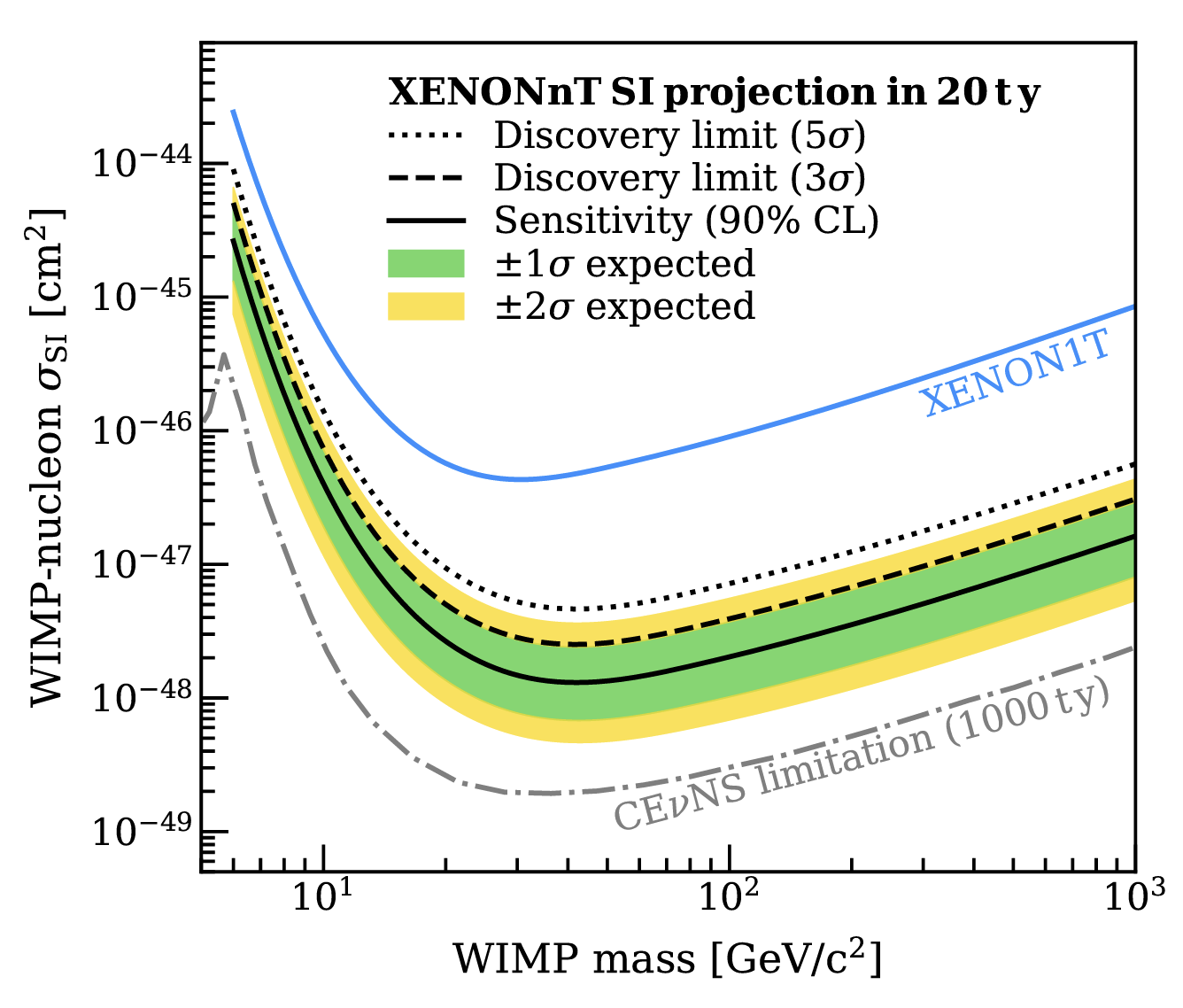}}
\end{center}
\caption{Reach of XENONnT with 5 years of data. From Ref.~\cite{XENON-2021}, used with permission.}
\label{XENON_reach}
\end{figure}
\begin{figure}[!tbp]
\begin{center}
\resizebox{1.00\columnwidth}{!}{\includegraphics{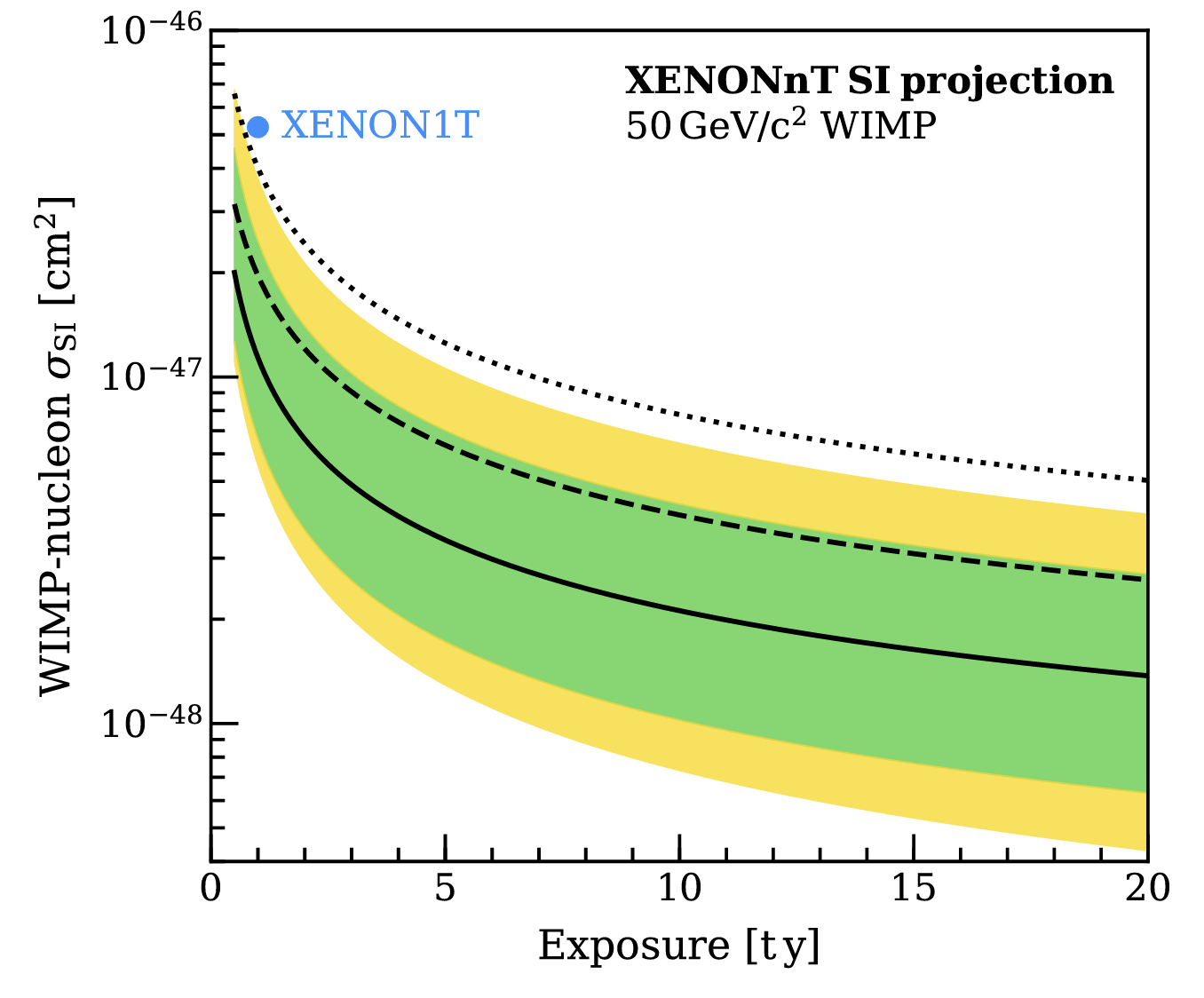}}
\end{center}
\caption{Reach of XENONnT for a 50 GeV WIMP in ton-years, with 4 tons fiducial mass. From Ref.~\cite{XENON-2021}, used with permission.}
\label{XENON_ty}
\end{figure}
\begin{figure}[!tbp]
\begin{center}
\resizebox{1.00\columnwidth}{!}{\includegraphics{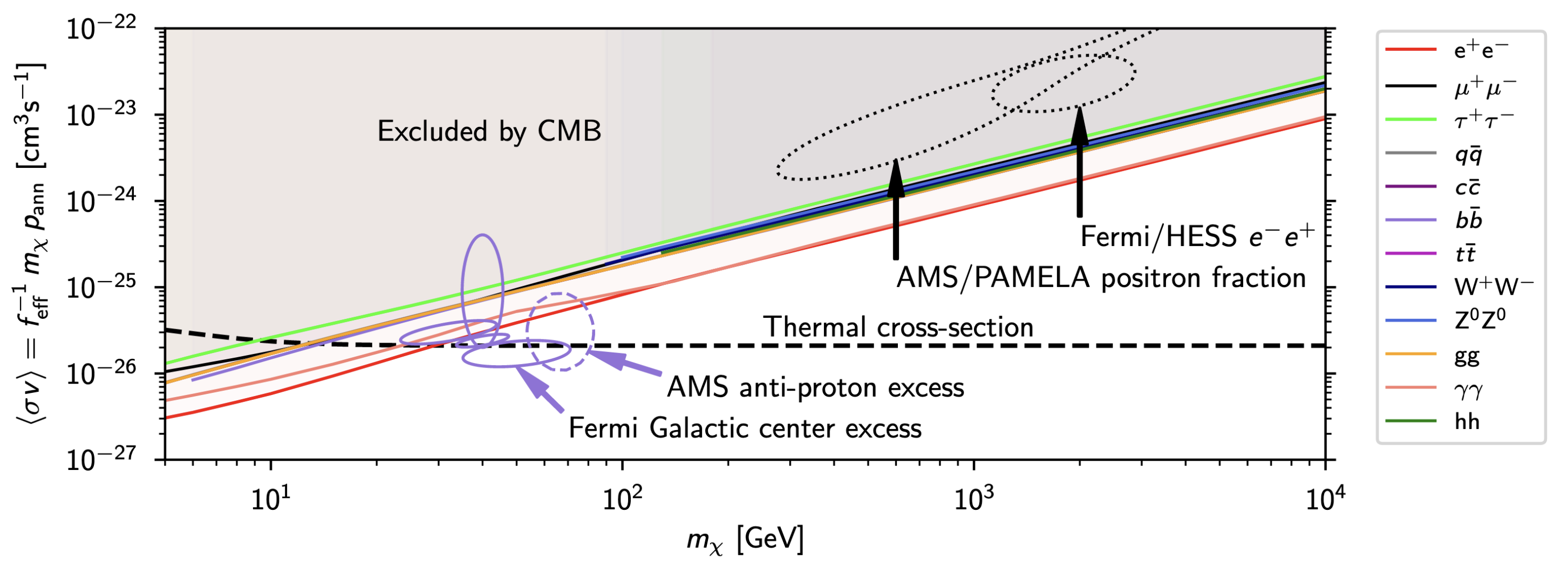}}
\end{center}
\caption{A high-mass dark matter candidate with high annihilation cross-section, which might have explained a positron excess observed by AMS-02 and other experiments, is excluded by the Planck data. The present candidate has a mass and cross-section consistent with this data. \\ Figure credit: Ref.~\cite{Planck}, ``Planck 2018 results. VI. Cosmological parameters'', Astronomy \&
Astrophysics manuscript, 10 August 2021, p. 61, Figure 46, reprinted with permission from ESO.}
\label{Planck}
\end{figure}
\begin{figure}[!tbp]
\begin{center}
\resizebox{0.65\columnwidth}{!}{\includegraphics{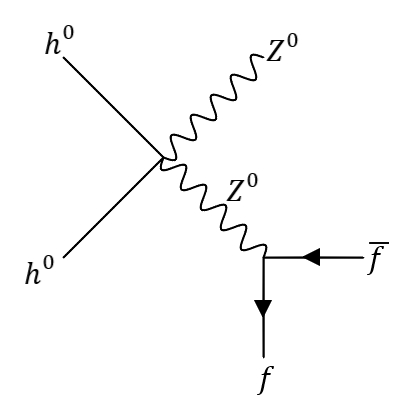}}
\end{center}
\caption{Representative diagram for annihilation of the present dark matter candidate via creation of Z bosons.}
\label{annZ}
\end{figure}
\begin{figure}[!tbp]
\begin{center}
\resizebox{0.65\columnwidth}{!}{\includegraphics{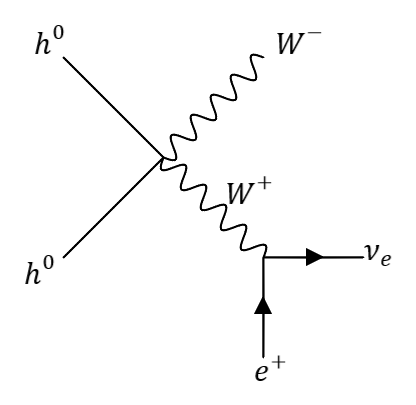}}
\end{center}
\caption{Representative diagram for annihilation of the present dark matter candidate via creation of W bosons.}
\label{annW}
\end{figure}
\begin{figure}[!tbp]
\begin{center}
\resizebox{0.65\columnwidth}{!}{\includegraphics{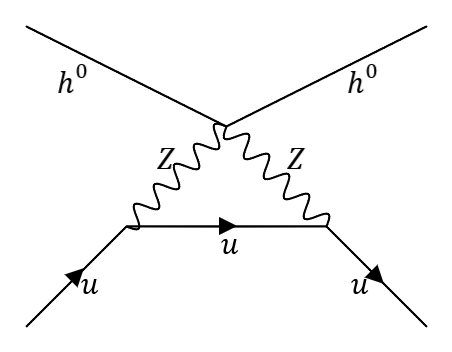}}
\end{center}
\caption{Representative diagram for direct detection of the present dark matter candidate with scattering via exchange of Z bosons.}
\label{dirZ}
\end{figure}
\begin{figure}[!tbp]
\begin{center}
\resizebox{0.65\columnwidth}{!}{\includegraphics{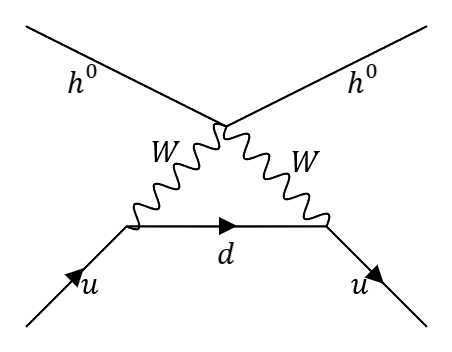}}
\end{center}
\caption{Representative diagram for direct detection of the present dark matter candidate with scattering via exchange of W bosons.}
\label{dirW}
\end{figure}
\begin{figure}[!tbp]
\begin{center}
\resizebox{0.65\columnwidth}{!}{\includegraphics{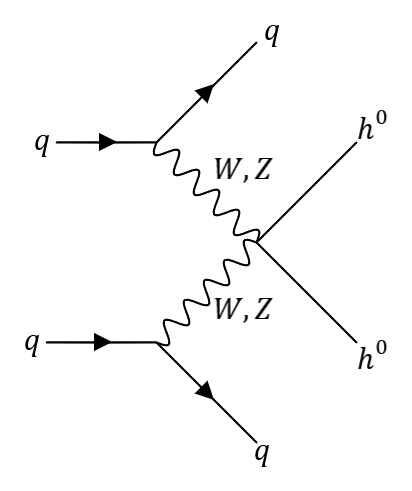}}
\end{center}
\caption{Representative diagram for collider detection of the present dark matter candidate via vector-boson fusion, with \\ $> 140$ GeV of missing energy accompanied by two jets.}
\label{col}
\end{figure}

The results above were obtained with MicrOMEGAs~\cite{MicrOMEGAs}. If we assume that the dark matter fraction $\Omega_{DM}$ is 0.27, that the present candidate constitutes all of the dark matter, and that the reduced Hubble constant $h$ is 0.73~\cite{Riess}, we obtain $\Omega_{DM} h^2= 0.144$. If it is instead assumed that a few percent of the dark matter consists of other components, making $\Omega_{DM} \approx 0.26$ for the present candidate, and that 
$h = 0.68$~\cite{Planck}, one obtains $\Omega_{DM} h^2 \approx 0.120$. (This value is equal to that obtained by Planck for all dark matter in an analysis that confirms the consistency of standard $\Lambda$CDM cosmology~\cite{Planck}.) Finally, as an extreme, we can consider 
$\Omega_{DM} = 0.22$ (for the present candidate) with $h = 0.68$, giving $\Omega_{DM} h^2= 0.102$.

Our calculations with MicrOMEGAs yield: \\ $\Omega_{DM} h^2= 0.162, 0.147, 0.134, 0.121, 0.098$ and \\
$\langle \sigma_{ann} v \rangle = 1.08, 1.19, 1.30, 1.43,  1.73 \times 10^{-26} $ cm$^3$/s, respectively, 
for $m_h =  69.5, 70.0, 70.5, 71.0, 72.0 $ GeV/c$^2$. 

We can conclude that $m_h = 70 - 72$ GeV/c$^2$ and that $\langle \sigma_{ann} v \rangle = 1.2 - 1.7 \times 10^{-26} $ cm$^3$/s. It is then reasonable to say that $m_h $ is about 70 GeV/c$^2$ and that correspondingly (with some bias toward the measured value of $h =0.73$ over the theoretical value of $h =0.68$ in the context of the present universe) $\langle \sigma_{ann} v \rangle \approx 1.2 \times 10^{-26} $ cm$^3$/s. 

It can be seen that our calculated $\langle \sigma_{ann} v \rangle $ with an approximately 70 GeV mass is well below the upper bounds of Fig.~\ref{Calore-2020} for any of the above values of $\Omega_{DM} h^2$.

Our calculated mass and $\langle \sigma_{ann} v \rangle $ are also consistent with analyses of the Galactic center gamma ray excess observed by Fermi-LAT~\cite{Leane-2,Goodenough,Fermi,Fermi-GCE,Leane-1,Leane} and the antiproton excess observed by AMS-02~\cite{Cuoco2,Cuoco,Cui,AMS-1,AMS-2}.

Ref.~\cite{Fermi-GCE} concludes that ``The center of the Milky Way is predicted to be the brightest region of $\gamma$-rays generated by self-annihilating dark matter particles. Excess emission about the Galactic center above predictions made for standard astrophysical processes has been observed in $\gamma$-ray data collected by the Fermi Large Area Telescope. It is well described by the square of a Navarro, Frenk, and White dark matter density distribution. Although other interpretations for the excess are plausible, the possibility that it arises from annihilating dark matter is valid.... Its spectral characteristics favor a dark matter particle with a mass in the range approximately from 50 to 190 (10 to 90) GeV and annihilation cross section approximately from $1 \times 10^{-26}$ to $4 \times 10^{-25}$ ($6 \times 10^{-27}$ to 
$2 \times 10^{-25}$) cm$^3$/s for pseudoscalar (vector) interactions.''

Ref.~\cite{AMS-1} finds that ``An excess of $\sim 10-20 $ GeV cosmic-ray antiprotons has been identified in the spectrum reported by the AMS-02 Collaboration.... After accounting for these uncertainties, we confirm the presence of a 4.7 $\sigma$ antiproton excess, consistent with that arising from a $m_{\chi} \approx 64-88$ GeV dark matter particle annihilating to $b \bar{b}$ with a cross section of $\sigma v = (0.8-5.2) \times 10^{-26} $ cm$^3$/s.''

Other analyses have yielded similar results, which are not very sensitive to the specific annihilation channel. 

At one time it may have appeared that a positron excess from AMS-02 and other experiments was evidence for a dominant dark matter particle at an energy of $\sim 800$ GeV or above. However, this interpretation has been ruled out by Planck~\cite{Planck-pos}, as shown in Fig.~\ref{Planck}, and the excess has been attributed to pulsars~\cite{Hooper-positron}. 

The present dark matter candidate, with a mass of about 70 GeV and roughly a thermal cross-section, is fully consistent with the observations and conclusions represented by Fig.~\ref{Planck}.

The present candidate results from an extended Higgs sector, which is an inevitable consequence of a broader fundamental theory~\cite{statistical,perspective}. This candidate is one member of a class of particles which we have called ``higgsons''~\cite{DM2021a,DM2021b,DM2022a}, represented by $h$, to distinguish them from Higgs bosons $H$ and higginos $\widetilde{h}$. 

We recall that there are three kinds of particles in the Standard Model. After the first spin 1/2 fermion was discovered in 1897 (by J.J. Thompson), and the first spin 1 gauge boson was postulated in 1905 (by Einstein), many surprises lay ahead with major extensions of these two sectors. It is reasonable that similar surprises and extensions may lie ahead after the 2012 discovery of a scalar boson (by the CMS and ATLAS collaborations). 

In the present theory, there are both complex scalar Higgs fields, having their standard interactions, and real scalar higgson fields, each of which interacts only with itself and gauge bosons, via second-order interactions like those of (\ref{A4}).

The lightest higgson $h^0$ is stable because of the form of the interaction in (\ref{A4}): It can radiate gauge bosons, annihilate into gauge bosons as in Figs.~\ref{annZ} and \ref{annW}, scatter via exchange of gauge bosons as in Figs.~\ref{dirZ} and \ref{dirW}, and be created in pairs as in Fig.~\ref{col}, but not decay, since a single initial $h^0$ implies a final state containing $h^0$ and two gauge bosons. 

With R-parity conserved, the lightest superpartner (LSP) is stable for a different reason, so the lightest higgson can coexist with the LSP (and with other particles stable for other reasons, such as axions~\cite{Sikivie,Tanner}). The present theory unavoidably predicts (broken) susy at some energy scale, and is compatible with well-motivated hypothetical particles such as axions.

However, the lightest higgson is assumed to be the dominant constituent, because it is difficult to reconcile the LSP in natural susy with the various experimental limits~\cite{Baer-Barger-2016,Baer-Barger-2018,Roszkowski-2018,Baer-Barger-2020,Tata-2020,Bertone,Peskin-2015,Olive1} and the other candidates tend not to have well-defined masses or couplings.

The present dark matter WIMP should be barely detectable by existing experiments, but certainly detectable with planned experiments such as DARWIN~\cite{DARWIN}. 

When a dark matter particle is discovered, rival claims to its nature can ultimately be determined by its properties (principally mass and interactions) and the general phenomenology associated with it. For example, various \textit{ad hoc} extended Higgs models tend to predict processes that do not exist in the present theory, with one example (the inert doublet model) discussed in detail in Ref.~\cite{DM2021a}.

To summarize the most important points: 

The present candidate is consistent with all current experimental and observational data.

The scattering processes of Figs.~\ref{dirZ} and \ref{dirW} lead to a cross-section for direct detection in Xe based experiments which we estimate to be slightly above $10^{-48}$ cm$^2$, placing it barely within reach of LZ and XENONnT within about the next 5 years, and definitely within reach of DARWIN. 

The creation processes of Fig.~\ref{col} lead to a cross-section for collider detection which we estimate to be $\sim 1$ femtobarn, which may place it barely within reach of the high-luminosity LHC within about 15 years, and definitely within reach of still more powerful colliders on a longer time scale. The signature in a proton collider is $> 140$ GeV of missing transverse energy with two quark jets. 

The annihilation processes of Figs.~\ref{annZ} and \ref{annW} have a cross\-section given by $\langle \sigma_{ann} v \rangle \approx 1.2 \times 10^{-26} $ cm$^3$/s. The mass and annihilation cross-section inferred in careful analyses of the gamma rays observed by Fermi-LAT and the antiprotons observed by AMS-02 are consistent with those calculated here, so indirect detection may already have been achieved.

\appendix

\section{\label{sec:appA}Action for scalar bosons and auxiliary fields}

In this appendix we quote some relevant results of Refs.~\cite{statistical} and \cite{perspective}, where the action for scalar boson fields has the form
\begin{eqnarray}
 S_{matter}  =\int d^{4}x\, e \, \overline{\mathcal{L}}_{scalar} 
 \label{A1}
 \end{eqnarray}
 \begin{align}
\overline{\mathcal{L}}_{scalar} =&
 \sum_R  \phi_R ^{\dagger } \left( x\right)    \left( D^{\mu}  D_{\mu } -  \frac{1}{4} R \right)  \phi_R \left( x\right)  
 + \sum_R F_R^{\dagger }\left( x\right) F_R\left( x\right)  \nonumber \\
&  +\sum_s \varphi _{s} \left(  \nabla^{\mu}\nabla_{\mu} - \frac{1}{4} R \right) \varphi _{s}  + \overline{\mathcal{L}}_{h-int}
\label{A2}
\end{align}
in a general coordinate system, but before masses and further interactions result from symmetry breakings and other effects.  
The $\phi_R$ are complex one-component Higgs fields, the $F_R$ are the one-component auxiliary fields of supersymmetry, and the $\varphi _{s}$ are real one-component higgson fields.
Each higgson field can be treated (and quantized) like a standard real scalar field, but with no quantum numbers and only second-order interactions.

Here 
\begin{align} 
 D_{\mu } = \nabla_{\mu} - i A_{\mu} 
\label{A3}
\end{align}
is the full covariant derivative, including the effects of both gravitational and gauge curvature, 
$R$ is the gravitational (Ricci) curvature scalar, and  $e = \left\vert \det \, e_{\mu }^{\alpha }\right\vert =\left( -\det \, g_{\mu \nu} \right) ^{1/2}$. The second-order gauge interactions of the higgson fields have been isolated in the last term, which can be written explicitly as 
\begin{align}
\overline{\mathcal{L}}_{h-int} =  \frac{g^{2}}{\left( 2 \cos \theta_W \right)^2}\, h_{s} \, Z^{\mu }Z_{\mu }\, h _{s} + \frac{g^{2}}{2} \, h _{s} \, W^{\mu +}W_{\mu }^{-} \, h _{s}
\label{A4}
\end{align}
in the electroweak sector, where it is assumed that there is no higgson condensate, so that $\varphi _{s} = h _{s}$, with the convention that $h_s$ is used to represent both a field and the particle which is an excitation of that field.

The higgson fields have only second-order interactions because they are the amplitude modes for Majorana-like bosonic fields that are constructed from primitive fields $\Phi_S$ and their charge conjugates $\Phi_S^{c}$:
\begin{align}
\Phi_S=\frac{1}{\sqrt{2}}\left( 
\begin{array}{c}
\Phi_S \\ 
\Phi_S^{c}
\end{array}
\right) \;.
\label{A5}
\end{align}
The first-order terms then cancel~\cite{DM2021a}. In addition, Yukawa couplings cannot exist and there is no mechanism for higgson-Higgs couplings. As a result, the cross-sections for annihilation, scattering, and creation are relatively small, making them consistent with current experimental and observational limits, while still within reach of experiments that have recently begun taking data or else are planned for the foreseeable future.


\begin{thebibliography}{99}

\bibitem{Silk}Gianfranco Bertone, Dan Hooper, and Joseph Silk, ``Particle dark matter: evidence, candidates and constraints'', Physics Reports 405, 279 (2005), arXiv:hep-ph/0404175.

\bibitem{Mambrini}Yann Mambrini, \textit{Particles in the Dark Universe: A Student’s Guide to Particle Physics and Cosmology} (Springer, 2021).

\bibitem{Snowmass}Daniel Green et al., ``Snowmass Theory Frontier: Astrophysics and Cosmology'', arXiv:2209.06854 [hep-ph].

\bibitem{pdg}L. Baudis and S. Profumo, ``Dark Matter'', in R.L. Workman et al. (Particle Data Group), Prog. Theor. Exp. Phys. 2022, 083C01 (2022), with updates at https://pdg.lbl.gov/.

\bibitem{DES-Wechsler}See e.g. T. M. C. Abbott et al. (DES Collaboration), ``Dark Energy Survey Year 3 Results: Constraints on extensions to CDM with weak lensing and galaxy clustering'', Phys. Rev. D 105, 023520 (2022), arXiv:2105.13549 [astro-ph.CO].

\bibitem{Strigari-2014}J. Billard, E. Figueroa-Feliciano, and L. Strigari, “Implication of neutrino backgrounds on the reach of next generation dark matter direct detection experiments”, Phys. Rev. D 89, 023524 (2014), arXiv:1307.5458 [hep-ph].

\bibitem{Calore-2020}Alexandre Alvarez, Francesca Calore, Anna Genina, Justin Read, Pasquale Dario Serpico, and Bryan Zaldivar, ``Dark matter constraints from dwarf galaxies with data-driven J-factors'',  JCAP 09, 004 (2020), arXiv:2002.01229 [astro-ph.HE]. 

\bibitem{collider}Antonio Boveia and Caterina Doglioni, ``Dark Matter Searches at Colliders'', Annu. Rev. Nucl. Part. Sci. 68, 429 (2018), arXiv:1810.12238 [hep-ex].

\bibitem{pdg2}O. Buchmuller and P. de Jon, ``Supersymmetry, Part II (Experiment)'', same pdg issue as for Ref.~\cite{pdg}.

\bibitem{DM2021a}Reagan Thornberry, Maxwell Throm, John Killough, Dylan Blend, Michael Erickson, Brian Sun, Brett Bays, Gabriel Frohaug, and Roland E. Allen, ``Experimental signatures of a new dark matter WIMP'', EPL [European Physics Letters] 134, 49001 (2021), arXiv:2104.11715 [hep-ph], and references therein.

\bibitem{DM2021b}Caden LaFontaine, Bailey Tallman, Spencer Ellis, Trevor Croteau, Brandon Torres, Sabrina Hernandez, Diego Cristancho Guerrero, Jessica Jaksik, Drue Lubanski, and Roland E. Allen, ``A Dark Matter WIMP That Can Be Detected and Definitively Identified with Currently Planned Experiments'', Universe 7, 270 (2021), arXiv:2107.14390 [hep-ph].

\bibitem{DM2022a}Bailey Tallman, Alexandra Boone, Caden LaFontaine, Trevor Croteau, Quinn Ballard, Sabrina Hernandez, Spencer Ellis, Adhithya Vijayakumar, Fiona Lopez, Samuel Apata, Jehu Martinez, and Roland Allen, ``Indirect detection, direct detection, and collider detection cross-sections for a 70 GeV dark matter WIMP'', Proceedings of Science (in press) [proceedings of the 41st International Conference on High Energy Physics, ICHEP 2022].

\bibitem{Steigman} G. Steigman, B. Dasgupta, and J. F. Beacom, ``Precise relic WIMP abundance and its impact on searches for dark matter annihilation'', Phys. Rev. D 86 023506 (2012), arXiv:1204.3622 [hep-ph].

\bibitem{susy-DM-1996}G. Jungman, M. Kamionkowski, and K. Griest, ``Supersymmetric Dark Matter'', Phys. Rept. 267, 195 (1996), arXiv:hep-ph/9506380.

\bibitem{Baer-Tata}H. Baer and X. Tata, \textit{Weak Scale Supersymmetry: From Superfields to Scattering Events} (Cambridge University Press, 2006), and references therein.

\bibitem{Kane} \textit{Perspectives on Supersymmetry II}, edited by G. L. Kane (World Scientific, 2010), and references therein.

\bibitem{snowmass-2013}Sebastian Arrenberg et al. [Snowmass 2013 CF4 Working Group Report], ``Dark Matter in the Coming Decade: Complementary Paths to Discovery and Beyond'',  arXiv:1310.8621 [hep-ph].

\bibitem{Baer-Roszkowski-2015}Howard Baer, Ki-Young Choi, Jihn E. Kim, and Leszek Roszkowski, ``Dark matter production in the early Universe: beyond the thermal WIMP paradigm'', Phys. Rep. 555, 1 (2015), arXiv:1407.0017 [hep-ph].

\bibitem{Baer-Barger-2016}Howard Baer, Vernon Barger, and Hasan Serce, ``SUSY under siege from direct and indirect WIMP detection experiments'', Phys. Rev. D 94, 115019 (2016), arXiv:1609.06735 [hep-ph].

\bibitem{Baer-Barger-2018}Howard Baer, Vernon Barger, Dibyashree Sengupta, and Xerxes Tata, ``Is natural higgsino-only dark matter excluded?'', Eur. Phys. J. C 78, 838 (2018), arXiv:1803.11210 [hep-ph].

\bibitem{Roszkowski-2018}Leszek Roszkowski, Enrico Maria Sessolo, and Sebastian Trojanowski, ``WIMP dark matter candidates and searches - current status and future prospects'', Rept. Prog. Phys. 81, 066201 (2018), arXiv:1707.06277 [hep-ph].

\bibitem{Baer-Barger-2020}Howard Baer, Dibyashree Sengupta, Shadman Salam, Kuver Sinha, and Vernon Barger, ``Midi-review: Status of weak scale supersymmetry after LHC Run 2 and ton-scale noble liquid WIMP searches", arXiv:2002.03013 [hep-ph].

\bibitem{Tata-2020}Xerxes Tata, ``Natural Supersymmetry: Status and Prospects'', arXiv:2002.04429 [hep-ph].

\bibitem{MicrOMEGAs}G. Belanger, F. Boudjema, A. Pukhov, and A. Semenov, ``micrOMEGAs: a tool for dark matter studies'',  Nuovo Cimento 33, 111 (2010), arXiv:1005.4133 [hep-ph], and references therein; http://lapth.cnrs.fr/micromegas/.

\bibitem{Riess}Adam G. Riess, Wenlong Yuan, Lucas M. Macri, Dan Scolnic, Dillon Brout, Stefano Casertano, David O. Jones, Yukei Murakami, Louise Breuval, Thomas G. Brink, Alexei V. Filippenko, Samantha Hoffmann, Saurabh W. Jha, W. D\'{a}rcy Kenworthy, Gagandeep Anand, John Mackenty, Benjamin E. Stahl, and Weikang Zheng, ``A Comprehensive Measurement of the Local Value of the Hubble Constant with 1 km/s/Mpc Uncertainty from the Hubble Space Telescope and the SH0ES Team'', ApJL 934, L7 (2022), arXiv:2112.04510 [astro-ph.CO].

\bibitem{Planck}Planck Collaboration, ``Planck 2018 results. VI. Cosmological parameters'', arXiv:1807.06209 [astro-ph.CO]. 

\bibitem{LZ-2022}J. Aalbers et al. [LUX-ZEPLIN (LZ) Collaboration], ``First Dark Matter Search Results from the LUX-ZEPLIN (LZ) Experiment'', arXiv:2207.03764 [hep-ex].

\bibitem{LZ-1000}Kelly Stifter [on behalf of the LZ collaboration], ``First dark matter search results from the LUX-ZEPLIN (LZ) experiment'', talk at International Conference on Neutrinos and Dark Matter (NuDM-2022), 
September 25-28, 2022, Sharm El-Sheikh, Egypt.

\bibitem{XENON-2021}E. Aprile et al., ``Projected WIMP Sensitivity of the XENONnT Dark Matter Experiment'', JCAP11, 031 (2020), 

\bibitem{Leane-2}Rebecca K. Leane, ``Indirect Detection of Dark Matter in
the Galaxy'', arXiv:2006.00513 [hep-ph].

\bibitem{Goodenough}Lisa Goodenough and Dan Hooper, ``Possible Evidence 
For Dark Matter Annihilation In The Inner Milky Way From The Fermi Gamma Ray 
Space Telescope'', arXiv:0910.2998 [hep-ph].

\bibitem{Fermi}Vincenzo Vitale and Aldo Morselli (for the Fermi/LAT Collaboration), 
``Indirect Search for Dark Matter from the center of the Milky Way with the Fermi-Large 
Area Telescope'', arXiv:0912.3828 [astro-ph.HE].

\bibitem{Fermi-GCE}Christopher Karwin, Simona Murgia, Tim M. P. Tait, Troy
A. Porter, and Philip Tanedo, ``Dark matter interpretation of the Fermi-LAT
observation toward the Galactic Center'', Phys. Rev. D 95, 103005 (2017),
arXiv:1612.05687 [hep-ph], and references therein.

\bibitem{Leane-1}Rebecca K. Leane, Tracy R. Slatyer, John F. Beacom, and
Kenny C. Y. Ng, ``GeV-scale thermal WIMPs: Not even slightly ruled out'',
Phys. Rev. D 98, 023016 (2018), arXiv:1805.10305 [hep-ph].

\bibitem{Leane}Rebecca K. Leane and Tracy R. Slatyer, ``Revival of the Dark
Matter Hypothesis for the Galactic Center Gamma-Ray Excess'', Phys. Rev.
Lett. 123, 241101 (2019), arXiv:1904.08430 [astro-ph.HE], and references
therein.
\bibitem{Cuoco2}Alessandro Cuoco, Jan Heisig, Michael Korsmeier, and Michael Kr\"{a}mer,
``Probing dark matter annihilation in the Galaxy with antiprotons and gamma rays'',
JCAP 10, 053 (2017), arXiv:1704.08258 [astro-ph.HE].

\bibitem{Cuoco}Alessandro Cuoco, Michael Kr\"{a}mer, and Michael Korsmeier, 
``Novel dark matter constraints from antiprotons in the light of AMS-02'', 
Phys. Rev. Lett. 118, 191102 (2017), arXiv:1610.03071 [astro-ph.HE].

\bibitem{Cui}Ming-Yang Cui, Qiang Yuan, Yue-Lin Sming Tsai, and Yi-Zhong Fan, 
``Possible dark matter annihilation signal in the AMS-02 antiproton data'',
Phys. Rev. Lett. 118, 191101 (2017), arXiv:1610.03840 [astro-ph.HE].

\bibitem{AMS-1}Ilias Cholis, Tim Linden, and Dan Hooper, ``A Robust Excess
in the Cosmic-Ray Antiproton Spectrum: Implications for Annihilating Dark
Matter'', Phys. Rev. D 99, 103026 (2019), arXiv:1903.02549 [astro-ph.HE].

\bibitem{AMS-2}Alessandro Cuoco, Jan Heisig, Lukas Klamt, Michael
Korsmeier, and Michael Kr\"{a}mer, ``Scrutinizing the evidence for dark
matter in cosmic-ray antiprotons'', Phys. Rev. D 99, 103014 (2019),
arXiv:1903.01472 [astro-ph.HE].

\bibitem{Planck-pos}See Fig. 46 of Ref.~\cite{Planck}.

\bibitem{Hooper-positron}Dan Hooper, Ilias Cholis, Tim Linden, and Ke Fang,
``HAWC Observations Strongly Favor Pulsar Interpretations of the Cosmic-Ray
Positron Excess'', Phys. Rev. D 96, 103013 (2017), arXiv:1702.08436
[astro-ph.HE].

\bibitem{statistical}R. E. Allen, ``Predictions of a fundamental statistical picture'', arXiv:1101.0586 [hep-th].

\bibitem{perspective}R. E. Allen, ``Some unresolved problems from a fresh perspective'', under review.

\bibitem{Sikivie}P. Sikivie, ``Experimental Tests of the ``Invisible'' Axion'', Phys. Rev. Lett. 51, 1415 (1983), and references therein.

\bibitem{Tanner}P. Sikivie, N. Sullivan, and D. B. Tanner, ``Proposal for Axion Dark Matter Detection Using an $LC$ Circuit'', Phys. Rev. Lett. 112,131301 (2014), arXiv:1310.8545 [hep-ph], and references therein.

\bibitem{Bertone} Gianfranco Bertone, ``The moment of truth for WIMP dark matter'', Nature 468, 389 (2010), arXiv:1011.3532 [astro-ph.CO].

\bibitem{Peskin-2015} Michael E. Peskin, ``Supersymmetric dark matter in the harsh light of the Large Hadron Collider'', Proc. Natl . Acad. Sci. USA 112, 12256 (2015).

\bibitem{Olive1} Keith A. Olive, ``Supersymmetric Dark Matter or Not'', proceedings of 11th International Workshop on Dark Side of the Universe, arXiv:1604.07336 [hep-ph].

\bibitem{DARWIN}J. Aalbers et al. (DARWIN collaboration), ``DARWIN: towards the ultimate dark matter detector'', JCAP 1611, 017 (2016) , arXiv:1606.07001 [astro-ph.IM].

\end{thebibliography}
\end{document}